\documentclass[pdflatex,sn-mathphys-num]{sn-jnl}

\usepackage{graphicx}%
\usepackage{multirow}%
\usepackage{amsmath,amssymb,amsfonts}%
\usepackage{amsthm}%
\usepackage{mathrsfs}%
\usepackage[title]{appendix}%
\usepackage{xcolor}%
\usepackage{textcomp}%
\usepackage{manyfoot}%
\usepackage{booktabs}%
\usepackage{algorithm}%
\usepackage{algorithmicx}%
\usepackage{algpseudocode}%
\usepackage{listings}%
\usepackage{booktabs} %
\usepackage[T1]{fontenc}

\theoremstyle{thmstyleone}%
\theoremstyle{thmstyletwo}%

\theoremstyle{thmstylethree}%
\usepackage{etoolbox}
\makeatletter
\def\aufullcenter{\leftskip=0pt plus 0.5fil\rightskip=0pt plus 0.5fil%
  \parfillskip=0pt\let\hb=\break}
\renewcommand{\Titlefont}{\reset@font\fontsize{17bp}{20bp}\selectfont\aufullcenter}
\renewcommand{\Authorfont}{\reset@font\fontsize{9bp}{10.5bp}\selectfont\aufullcenter}
\renewcommand{\addressfont}{\reset@font\fontsize{8bp}{9.2bp}\selectfont\aufullcenter}

\patchcmd{\@maketitle}{\vskip21pt}{\vskip0pt}{}{\@latex@warning{TOPSKIP not patched}}
\patchcmd{\@maketitle}{\removelastskip\vskip20pt\nointerlineskip}{\nointerlineskip}{}{\@latex@warning{TITLESKIP not patched}}
\patchcmd{\@maketitle}{\artauthors\par}
  {\artauthors\par\vskip2pt{\Authorfont(IceCube Collaboration)\par}}
  {}{\@latex@warning{collaboration line not patched}}
\makeatother

\begin{document}

\title[Article Title]{Planck-Scale Constraints on Quantum Gravity-Induced Neutrino Decoherence with IceCube
}

\author[16]{\fnm{R.} \sur{Abbasi}}
\author[63]{\fnm{M.} \sur{Ackermann}}
\author[17]{\fnm{J.} \sur{Adams}}
\author[10]{\fnm{J. A.} \sur{Aguilar}}
\author[21]{\fnm{M.} \sur{Ahlers}}
\author[22]{\fnm{J.M.} \sur{Alameddine}}
\author[35]{\fnm{S.} \sur{Ali}}
\author[43]{\fnm{N. M.} \sur{Amin}}
\author[41]{\fnm{K.} \sur{Andeen}}
\author[13]{\fnm{C.} \sur{Argüelles}}
% ...
\affil[10]{Université Libre de Bruxelles, Science Faculty CP230, B-1050 Brussels, Belgium}
% ...
\affil[13]{Department of Physics and Laboratory for Particle Physics and Cosmology, Harvard University, Cambridge, MA 02138, USA}
% ...
\affil[16]{Department of Physics, Loyola University Chicago, Chicago, IL 60660, USA}
\affil[17]{Dept. of Physics and Astronomy, University of Canterbury, Private Bag 4800, Christchurch, New Zealand}
% ...
\affil[21]{Niels Bohr Institute, University of Copenhagen, DK-2100 Copenhagen, Denmark}
\affil[22]{Dept. of Physics, TU Dortmund University, D-44221 Dortmund, Germany}
% ...
\affil[35]{Dept. of Physics and Astronomy, University of Kansas, Lawrence, KS 66045, USA}
% ...
\affil[41]{Department of Physics, Marquette University, Milwaukee, WI 53201, USA}
% ...
\affil[43]{Bartol Research Institute and Dept. of Physics and Astronomy, University of Delaware, Newark, DE 19716, USA}
% ...
\affil[63]{Deutsches Elektronen-Synchrotron DESY, Platanenallee 6, D-15738 Zeuthen, Germany}

\footnotetext[1]{also at Department of Space, Earth and Environment, Chalmers University of Technology, 412 96 Gothenburg, Sweden}
\footnotetext[2]{also at INFN Padova, I-35131 Padova, Italy}
\footnotetext[3]{also at Earthquake Research Institute, University of Tokyo, Bunkyo, Tokyo 113-0032, Japan}
\footnotetext[4]{now at INFN Padova, I-35131 Padova, Italy}


\abstract{\unboldmath  Many quantum gravity theories predict spacetime fluctuations at the Planck scale. Neutrinos, which interact only via the weak force and gravity, propagate over long distances without losing quantum coherence. Planck-scale fluctuations would spoil that coherence, damping the flavor oscillation probability over long baselines. This makes neutrinos sensitive interferometers for Planck-scale physics. We present a search for quantum decoherence using 368,071 atmospheric neutrinos detected by IceCube in the energy range 0.5-100~TeV, an order of magnitude higher in energy than any previous decoherence search. We find no evidence for decoherence and set the strongest constraints to date on this class of neutrino decoherence models, improving on previous limits by over an order of magnitude for decoherence rate scaling as $E^2$ and $E^3$. These results exclude natural Planck-suppressed decoherence for all energy scalings $E^n$ with $0\leq n \leq 3$, the first such exclusion at the Planck scale for $n=3$. 
}

\maketitle

\section{Introduction}\label{sec1}
When a quantum system interacts with an environment, its coherent superpositions are progressively degraded into classical mixtures. This process is known as decoherence and is described within the framework of open quantum systems~\cite{breuer2002}. Whether spacetime itself acts as a decoherent medium for propagating quantum states remains an open question in fundamental physics. A consistent quantum theory of gravity does not yet exist. However, model candidates generally predict that the smooth classical geometry of spacetime breaks down at the Planck scale, giving way to an underlying quantum structure of irreducible fluctuations~\cite{carlip2023}. Even if these effects are individually negligible at measurable energies, their cumulative influence on a propagating quantum state over macroscopic distances may generate observable departures from isolated, coherent evolution.

Gravitationally-induced decoherence has been proposed in a variety of contexts, including gravity-driven wavefunction collapse~\cite{penrose1996, diosi1987}, spacetime foam and virtual black hole models~\cite{hawking}, fluctuating minimal lengths~\cite{petruzziello2021}, and quantum spacetime-induced stochastic phase perturbations~\cite{arzano2023}. These models generically predict small departures from perfectly coherent quantum evolution.

Experimental searches for gravity-driven decoherence span multiple physical systems, including matter-wave interferometry and underground radiation detectors, both probing wavefunction collapse models~\cite{donadi2021, arnquist2022}. However, most laboratory tests probe microscopic length scales and energies far below those relevant for Planck-scale physics~\cite{addazi2022}.

Neutrino flavor oscillations provide a complementary and uniquely sensitive regime. As electrically neutral particles interacting only via the weak force and gravity, neutrinos propagate largely undisturbed over macroscopic lengths. So, even tiny decoherence effects accumulate into observable distortions of the flavor evolution.  Planck-suppressed effects are expected to grow with neutrino energy~\cite{stuttard_qg}, while standard oscillations depend only on the ratio of distance to energy and are largest at low energies. The scale of the sensitivities is therefore set by the experiments' baselines and energies rather than by how well they resolve oscillations. Accelerator and reactor experiments measure the oscillation pattern precisely, but their GeV-scale energies and shorter baselines limit their reach~\cite{JHEP09_2023_097}. Astrophysical neutrinos reach the highest energies and longest baselines, but the oscillation pattern is washed out by their large and unknown source distances, broad energy spectra, and finite detector resolution. The resulting oscillation-averaged flavor composition matches the asymptotic limit of decoherence, so it is indistinguishable from the standard expectation~\cite{stuttard_qg}. Atmospheric neutrinos traverse distances up to the Earth's diameter and retain coherence at multi-TeV energies~\cite{jung2001}. Combined with large statistics and a well-understood production mechanism, this makes them the most sensitive available probe of Planck-scale decoherence.

Decoherence damps flavor oscillations as a neutrino propagates, driving the flavor composition toward a constant value. To describe decoherence quantitatively, we use the open quantum system framework.  The most general Markovian, positive, trace-preserving master equation for an open quantum system is the Gorini-Kossakowski-Sudarshan-Lindblad (GKSL) equation, derived independently by Lindblad~\cite{lindblad} and by Gorini, Kossakowski, and Sudarshan~\cite{GKS1976} in 1976. Applied to the neutrino density matrix, this framework accommodates environment-induced decoherence through additional dissipative superoperators appended to the standard oscillation Hamiltonian, while preserving complete positivity and the conservation of total probability~\cite{benatti2000}. The energy dependence and flavor structure of these Lindblad operators are left as free phenomenological parameters, to be constrained by data and subsequently interpreted in terms of specific quantum gravity models. Previous experimental searches for such effects in the neutrino sector have used data from Super-Kamiokande~\cite{superk_deco}, MINOS and T2K~\cite{T2K_minos}, KamLAND, NOvA, RENO, and Daya Bay~\cite{JHEP09_2023_097}, KM3NeT~\cite{km3net_deco}, and IceCube~\cite{gp_deco_paper, coloma2018}. Experiments at lower neutrino energies place stronger constraints on decoherence that falls with energy. High-energy atmospheric neutrinos are instead the most sensitive probe of effects that are constant or grow with energy, which are more commonly motivated by quantum-gravity models.

The IceCube Neutrino Observatory is particularly well-suited for this search. IceCube consists of $5,160$ photodetectors called digital optical modules (DOMs)~\cite{DOM_paper} that instrument a cubic kilometer of Antarctic glacial ice at the South Pole. It detects Cherenkov radiation from charged particles produced in neutrino interactions. Charged-current interactions of muon neutrinos produce long-ranging muons that yield track-like signatures. Other interaction channels, including charged-current interactions of electron and tau neutrinos and neutral-current interactions of all flavors, produce more localized electromagnetic or hadronic particle showers (cascades)~\cite{ic_detector, cascade_ref}.

In this work, we present a search for neutrino decoherence using 10.7 years of upward-going atmospheric muon-neutrino data collected by the IceCube Neutrino Observatory, previously used in the search for eV-scale sterile neutrinos~\cite{alfonso_sterile}. The sample spans baselines up to the Earth's diameter and reconstructed energies from 0.5 to 100~TeV, an order of magnitude higher than the previous IceCube decoherence search~\cite{gp_deco_paper}. Together with an expanded dataset, refined event selection, and improved treatment of detector, flux, and cross-section systematics~\cite{methods&stability}, this extended energy reach improves the limits on energy-dependent decoherence by more than an order of magnitude.

\section{Quantum Decoherence}
\begin{figure}[ht]
    \centering
    \includegraphics[width=0.99\linewidth]{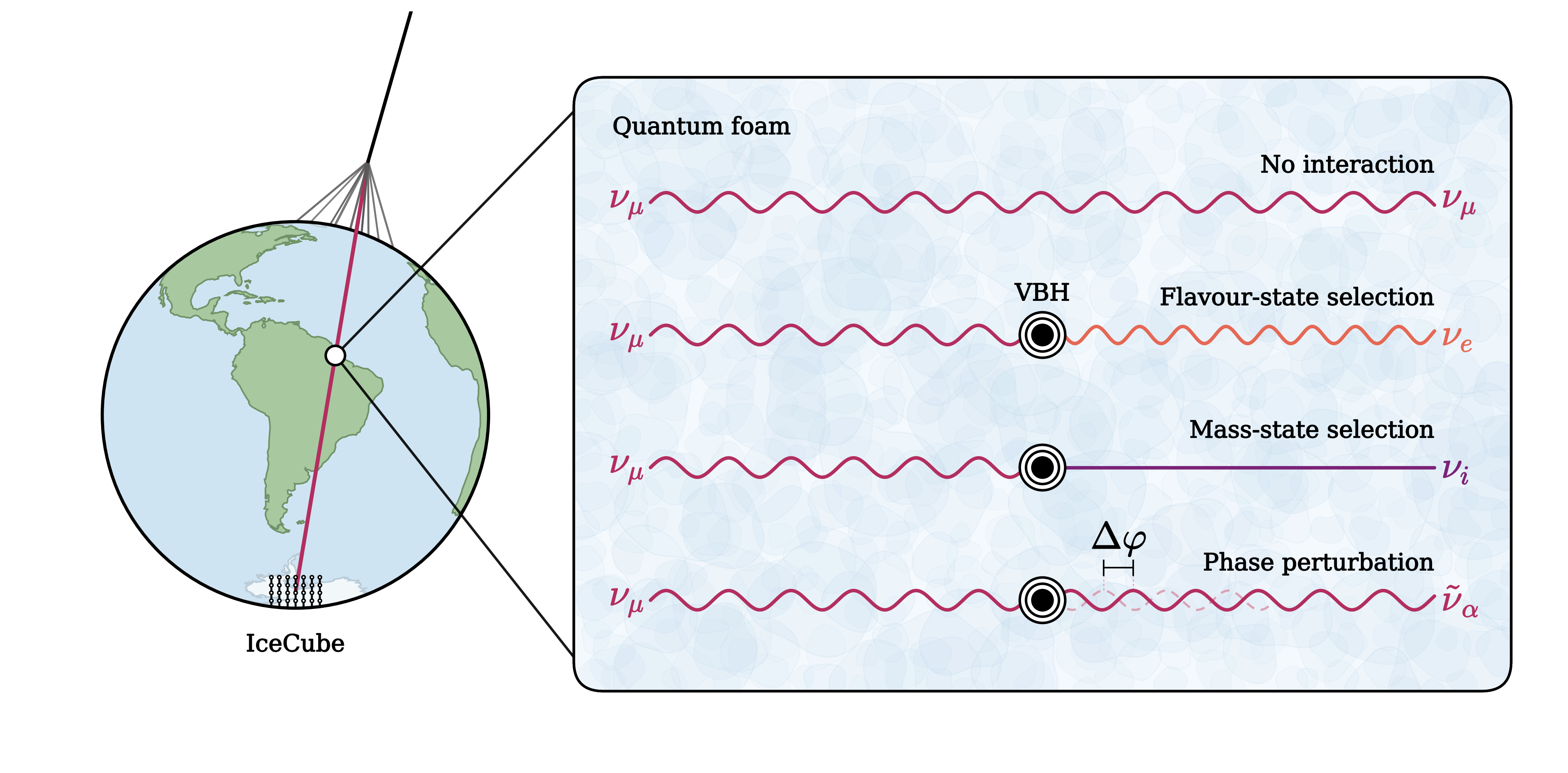}
    \caption{Schematic of neutrino decoherence from interactions with virtual black holes (VBHs). An atmospheric neutrino produced in a cosmic-ray air shower traverses the Earth to IceCube, passing through a fluctuating spacetime background. A $\nu_\mu$ that undergoes no interaction (top) propagates coherently. In the state selection model, the VBH interaction projects the neutrino onto a definite flavor state (second) or mass state (third). Both cases are described by the same decoherence operator and are treated together in this work. In the phase perturbation model (bottom), the interaction imparts a random phase shift $\Delta\varphi$ between mass states. In both models, averaging over the random fluctuations damps the observed oscillation pattern over long baselines.}
    \label{fig:schematic}
\end{figure}
\begin{figure}[ht]
    \centering
    \includegraphics[width=0.5\linewidth]{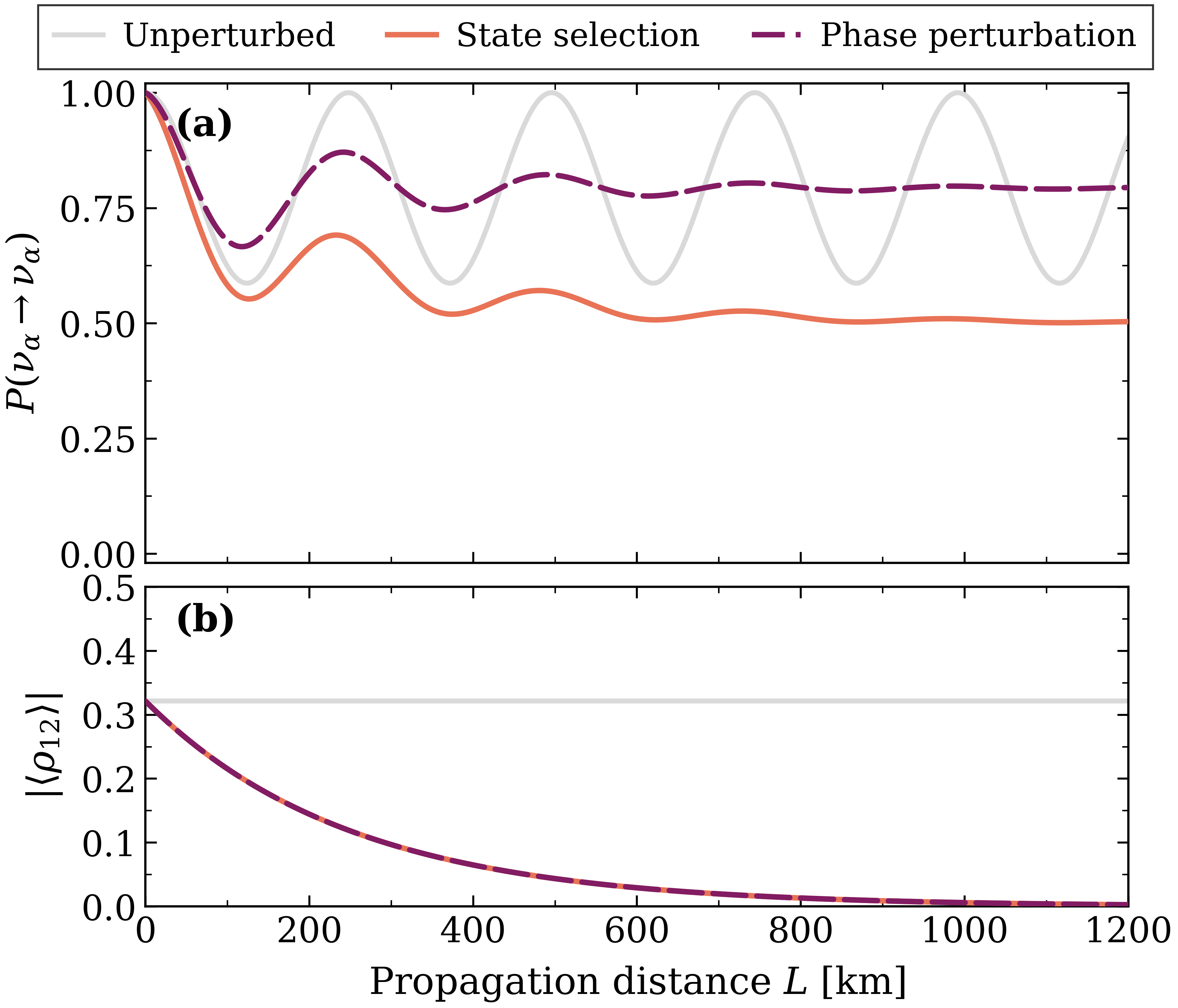}
    \caption{(a) Flavor survival probability $P(\nu_\alpha\to\nu_\alpha)$ as a function of propagation distance $L$ for a simplified 2-flavor model for the state selection and phase perturbation modes. (b) Coherence amplitude $|\langle\rho_{12}\rangle|$ in the mass basis. Both models produce exponential damping of coherence with distance due to stochastic interactions with VBHs, while producing different asymptotic flavor compositions. The full phenomenology of this model is described in Ref.~\cite{stuttard_qg}.}
    \label{fig:model}
\end{figure}

Atmospheric neutrino oscillations are well established and well-described by unitary quantum evolution~\cite{superK, deepCore}, since neutrinos interact only very weakly with matter. Thus, neutrino wavefunctions can propagate coherently over large distances. A common expectation across quantum-gravity approaches is that spacetime at the Planck length is not smooth but fluctuating, sometimes called spacetime foam. One concrete realization is a background of virtual black holes (VBHs): microscopic, transient black holes that form and evaporate on Planck timescales. A neutrino propagating through such a background scatters off them stochastically, and each interaction can alter the neutrino state irreversibly~\cite{hawking}. We study two different heuristic models for these interactions, illustrated in~\autoref{fig:schematic}: democratic mass/flavor state selection and mass state phase perturbation~\cite{stuttard_qg}. We will refer to these two interaction models as \textit{state selection} and \textit{phase perturbation}, respectively. The effects of these two models on an ensemble of propagating neutrinos are illustrated in~\autoref{fig:model} using a simplified two-flavor scenario, while the analysis itself is performed in the full three-flavor framework. State selection randomizes the neutrino's flavor or mass state, driving the ensemble towards flavor equilibrium, while phase perturbation preserves the state composition but introduces stochastic phase shifts that wash out coherent oscillation patterns over long distances.
We use the formalism of open quantum systems to encode the time evolution of a neutrino coupled to its environment, using the Lindblad master equation (in natural units with $\hbar=c=1$):
\begin{equation}
\dot{\rho}=-i[H,\rho]-\mathcal{D}[\rho],
\end{equation}
where $\rho$ is the density matrix, $H$ is the Hamiltonian of the neutrino system, and $\mathcal{D}[\rho]$ is the decoherence superoperator~\cite{lindblad}. The decoherence superoperator is dependent on the underlying physics causing the decoherence effect. For a three-neutrino system, it can be expanded in the basis of the Gell-Mann matrices ($b^\mu$) \cite{Georgi1999}: 
\begin{equation}
\mathcal{D}[\rho]=(D_{\mu\nu}\rho^{\nu})b^{\mu},
\end{equation} 
where $\rho^{\nu}$ are the coefficients of the system's density matrix expanded in the basis of $b^{\mu}$. The matrix $D_{\mu\nu}$ is a $9\times 9$ matrix that parametrizes the decoherence effects in a chosen model. The state selection model, in the long-distance limit, predicts a flux that is equally weighted in all neutrino flavors, independent of the basis in which the neutrino state is randomly projected. The phase perturbation model, in the long-distance limit, predicts a flux that tends toward an incoherent sum of mass eigenstates, reflecting the loss of phase coherence while preserving the mass-state populations. The simplest way to encode these averaging behaviors is by taking $D$ as diagonal matrices parametrized by a single real parameter $\Gamma$:
\begin{equation}
    D_{\text{state selection}} = \text{diag}(0, \Gamma, \Gamma, \Gamma, \Gamma, \Gamma, \Gamma,\Gamma, \Gamma),
\end{equation}
\begin{equation}
D_{\text{phase perturbation}} = \text{diag}(0, \Gamma, \Gamma, 0, \Gamma, \Gamma, \Gamma,\Gamma, 0).
\end{equation}
We assume that $\Gamma$ has a power-law energy dependence:
\begin{equation}
    \Gamma(E) = \Gamma_0\left(\frac{E}{E_0}\right)^n,
\end{equation}
where $\Gamma_0$ corresponds to the decoherence rate at the reference energy $E_0$ and $n$ denotes the energy dependence of the decoherence rate. In this analysis we will fix $E_0=1\,\text{TeV}$, leaving two free parameters: $n$ and $\Gamma_0$. Three-flavor neutrino propagation through the Earth is calculated using a modified version of \texttt{nuSQuIDS}~\cite{nusquids}, incorporating standard oscillations, Earth attenuation, $\nu_\tau$ regeneration, and Lindblad decoherence. 

\begin{figure}[t]
    \centering
    \includegraphics[width=0.99\linewidth]{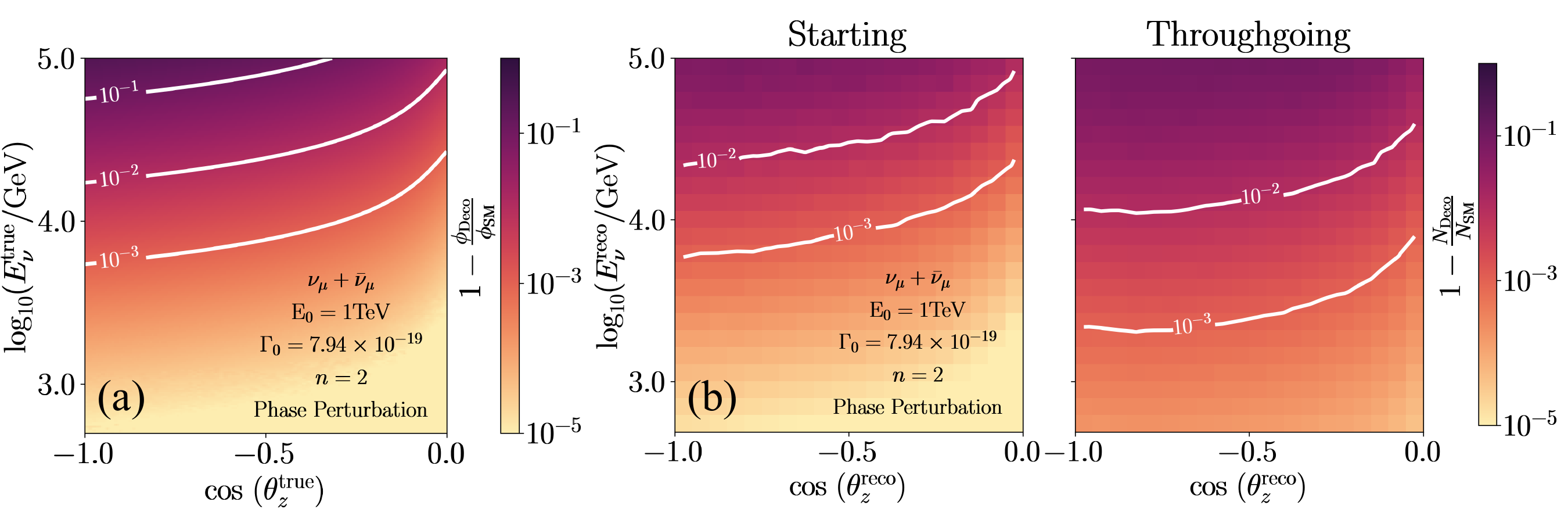}
    \caption{Fractional deviation of the phase perturbation model prediction for a representative decoherence scenario ($n=2$) to the best-fit standard three-flavor oscillation prediction. Panel (a) shows $1-\phi_{\text{Deco}}/\phi_{\text{SM}}$ for the $\nu_\mu$ flux as a function of true neutrino energy and zenith angle; panel (b) shows $1-N_{\text{Deco}}/N_{\text{SM}}$ for the expected event counts as a function of reconstructed energy and zenith angle for starting and throughgoing tracks respectively.}
    \label{fig:osc}
    
\end{figure}

An example oscillogram for a representative set of parameters is shown in \autoref{fig:osc}, showing the change in $\nu_\mu$ flux and rate as a function of energy and zenith angle across the high-energy IceCube $\nu_\mu$ sample. The data and simulation are binned in two dimensions: energy and the cosine of the zenith angle, as shown in \autoref{fig:osc}~(b). Because IceCube does not distinguish neutrinos from antineutrinos on an event-by-event basis, the sample primarily constrains their combined behavior and has limited sensitivity to potential differences between their decoherence parameters. We therefore restrict this analysis to CPT-conserving decoherence. 

We search for decoherence by fitting the reconstructed energy and zenith distributions of upward-going muon-neutrino events under the state-selection and phase-perturbation models described above. For each model, we test energy dependences $\Gamma(E)=\Gamma_0(E/E_0)^n$ with $n=0,1,2,3$, profiling over detector, flux, and interaction uncertainties as nuisance parameters. These values are chosen because Planck-suppressed effects are expected to become more pronounced at higher energies and IceCube is therefore most competitive for positive power-law indices. An $E^2$ dependence appears in several string-theory calculations of decoherence, both for propagation in a four-dimensional black hole background and for D-brane foam~\cite{ellis_4d_blackhole, ellis_dfoam}, making $n=2$ the best theoretically motivated of the cases we test. We also test $n=3$, the steepest scaling to which atmospheric neutrinos retain sensitivity. For $n>3$, a natural Planck-scale predicts coherence lengths longer than the Earth's diameter at TeV energies, so there is no sensitivity to be gained from atmospheric neutrinos~\cite{stuttard_qg}.

The analysis performs a simultaneous fit to the starting- and throughgoing-track samples (events with vertices inside and outside the instrumented detector volume, respectively). The resulting profile-likelihood scans over $\Gamma_0$ are used to set upper limits on decoherence strength at 90\% confidence-level (CL). Further details of the event selection, binning, nuisance parameters, and validation procedure are given in \nameref{sec4}.

\section{Results}\label{sec2}
This analysis constrains neutrino-VBH interaction models with 10.7 years of IceCube data using $368,071 $ up-going $\nu_\mu$ events with reconstructed energies ranging from $0.5$~TeV to $100$~TeV. The reconstruction and event selection are described in further detail in the \nameref{sec4} section and in Ref.~\cite{methods&stability}, which provides the detailed methodology accompanying Ref.~\cite{alfonso_sterile}. The same reconstruction and event selection are used in both Ref.~\cite{alfonso_sterile} and this work.

\begin{figure}
 \centering
    \includegraphics[width=0.75\linewidth]{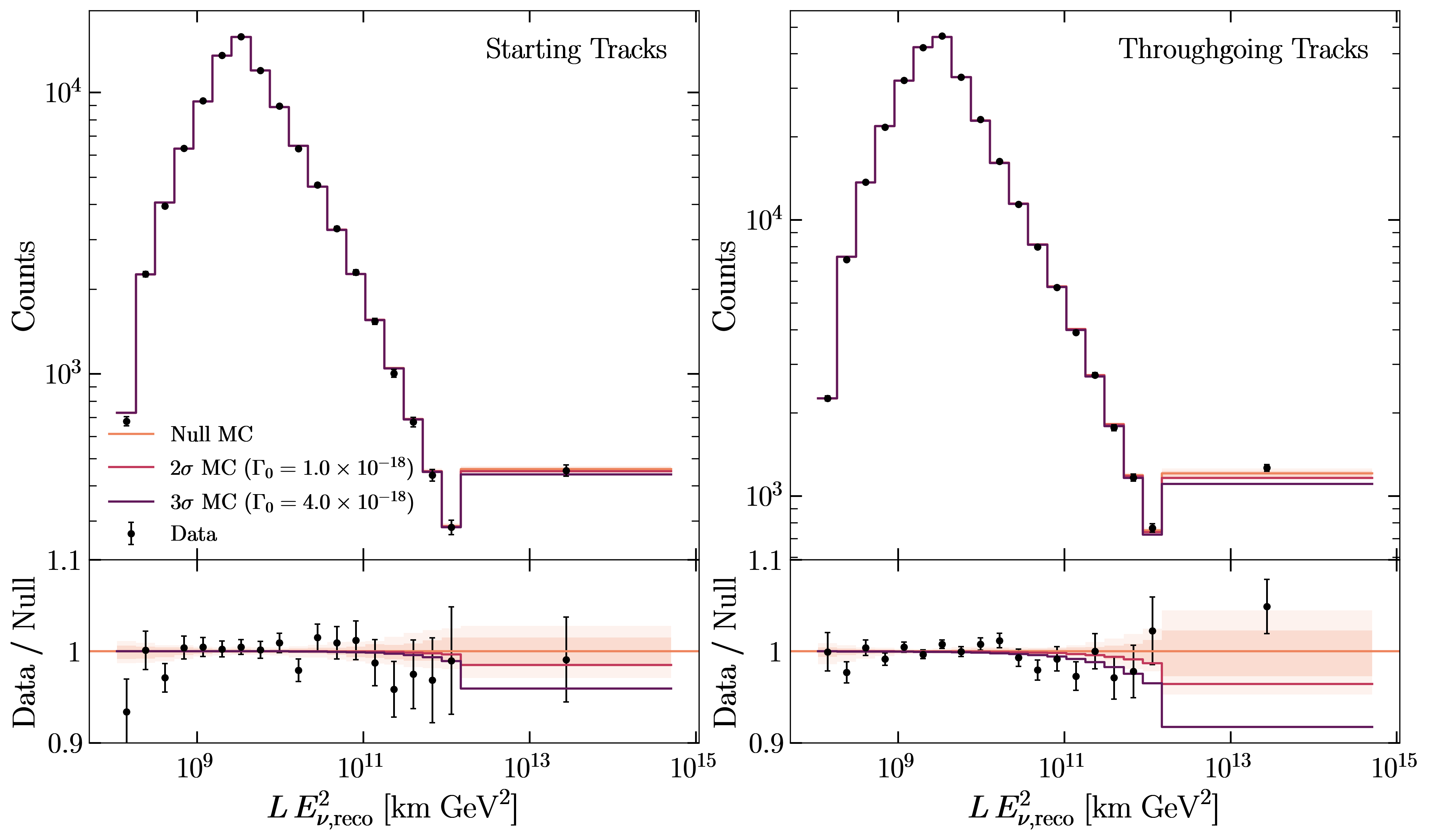}
    \caption{Data vs Monte Carlo as a function of $LE^n$ for the n=2, phase perturbation model for starting tracks (left) and throughgoing tracks (right). The length $L$ corresponds to the propagation baseline from the production point in the atmosphere to the detector. Black points show data with Poisson statistical errors. The orange line shows null Monte Carlo (expected counts under standard three-flavor oscillation), while the red and purple lines show Monte Carlo with decoherence strength ruled out at $2\sigma$ and $3\sigma$ by this analysis, respectively. The orange bands show 68\% and 95\% systematic uncertainties from posterior sampling of the nuisance parameters. The top plots show the comparison in counts, and the bottom plots show the ratio to the null MC. Note that this is not the analysis binning. }
    \label{fig:data_MC}
\end{figure}

 Though the data are binned in energy and zenith angle, \autoref{fig:data_MC} shows the data and expectations from Monte Carlo (MC) as a function of $LE^n$, where $L$ is the neutrino propagation distance from the production point in the atmosphere to the detector, calculated from the reconstructed zenith angle assuming a production altitude of 15~km. This combination controls the damping of the oscillation probability by decoherence. For the $n=2$ model shown, the predicted deficit is concentrated in the high-energy tail of the sample, while for $n=0$ the damping depends on baseline alone and the most vertical events dominate at all energies. The data track the null expectation across the full range of $LE^n$ , with no deficit where decoherence would be strongest.  The same is true for every model tested, as reflected in the limits below.
 
Since no evidence of decoherence is found, we set $90\%$ CL upper limits on the decoherence strength for each of the models we fit. The limits from this analysis can be seen in \autoref{tab:limits}. These constraints substantially narrow the viable parameter space for Planck-suppressed open-system extensions of neutrino oscillations across all decoherence models accessible to atmospheric neutrino baselines and energies.

\begin{table}[htbp]
\centering
\begin{tabular}{ccc}
\toprule
$n$ & State Selection $\Gamma_{90}$ (eV) & Phase Perturbation $\Gamma_{90}$ (eV) \\
\midrule
0 & $2.36 \times 10^{-15}$ & $2.41 \times 10^{-15}$ \\
1 & $4.58 \times 10^{-16}$ & $4.80 \times 10^{-16}$ \\
2 & $7.26 \times 10^{-19}$ & $7.26 \times 10^{-19}$ \\
3 & $6.05 \times 10^{-21}$ & $6.37 \times 10^{-21}$ \\
\bottomrule
\end{tabular}
\caption{90\% CL upper limits obtained from this analysis, with $E_0 =1\,\mathrm{TeV}$.}
\label{tab:limits}
\end{table}

\begin{figure}
    \centering
    \includegraphics[width=0.7\linewidth]{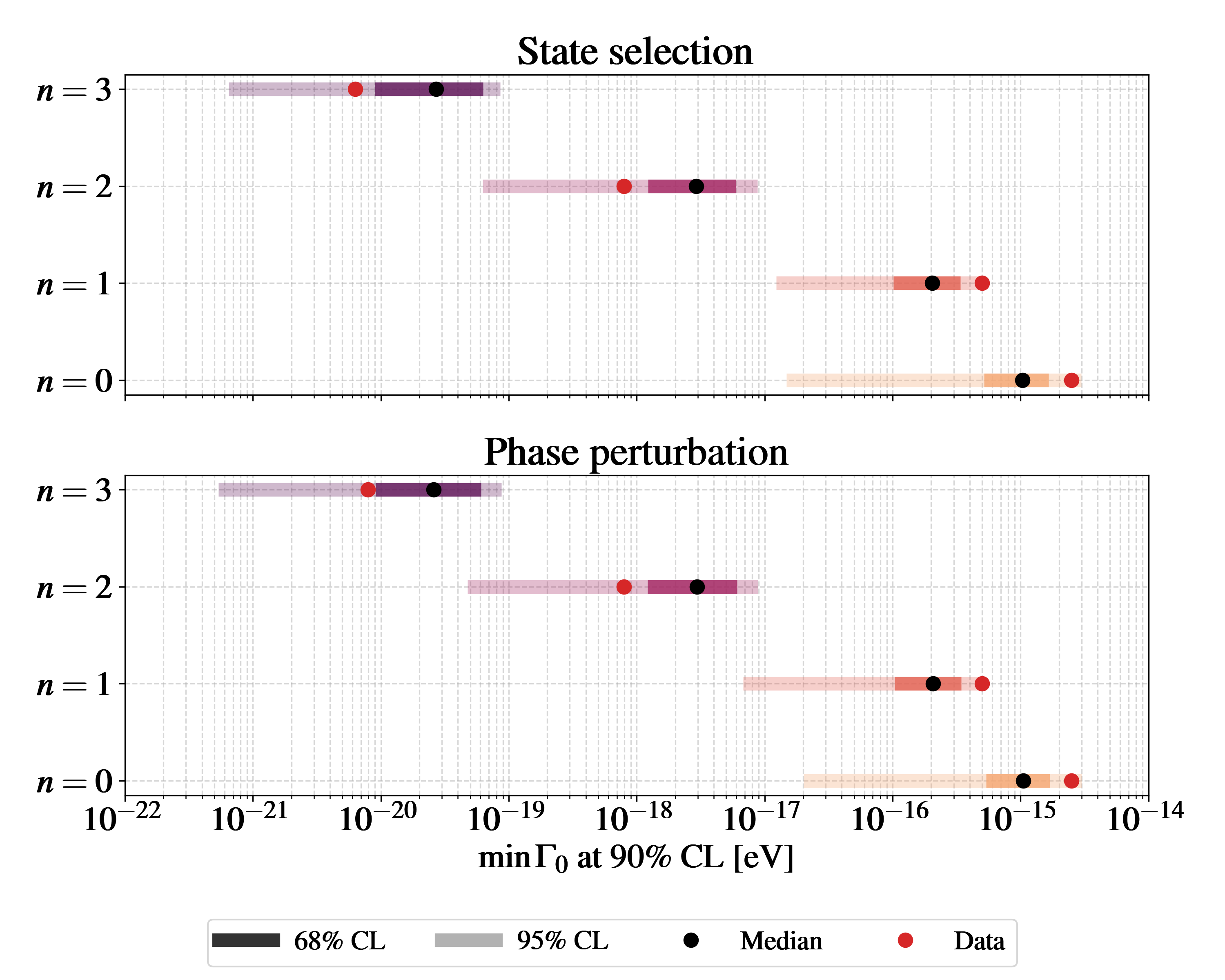}
    \caption{90\% CL upper limits (red) and sensitivity (black) for state selection (top) and phase perturbation (bottom) for each model of $n=0, 1, 2, 3$, with $E_0=1\,\text{TeV}$. Frequentist regions for 68\% CL (darker shaded region) and 95\% CL (lighter shaded region) of 1000 pseudoexperiment trials are overlaid.}
    \label{fig:brazil}
\end{figure}

We also show in \autoref{fig:brazil} the expected sensitivity ranges with $68\%$ and $95\%$ CL bands, which demonstrate that the upper limits we obtained are within the expected statistical fluctuations, compared to the upper limits from 1,000 pseudo-experiments from MC following the null hypothesis. The data fall within the $2\sigma$ CL bands for each model.

\begin{figure}
    \centering
    \includegraphics[width=0.9\linewidth]{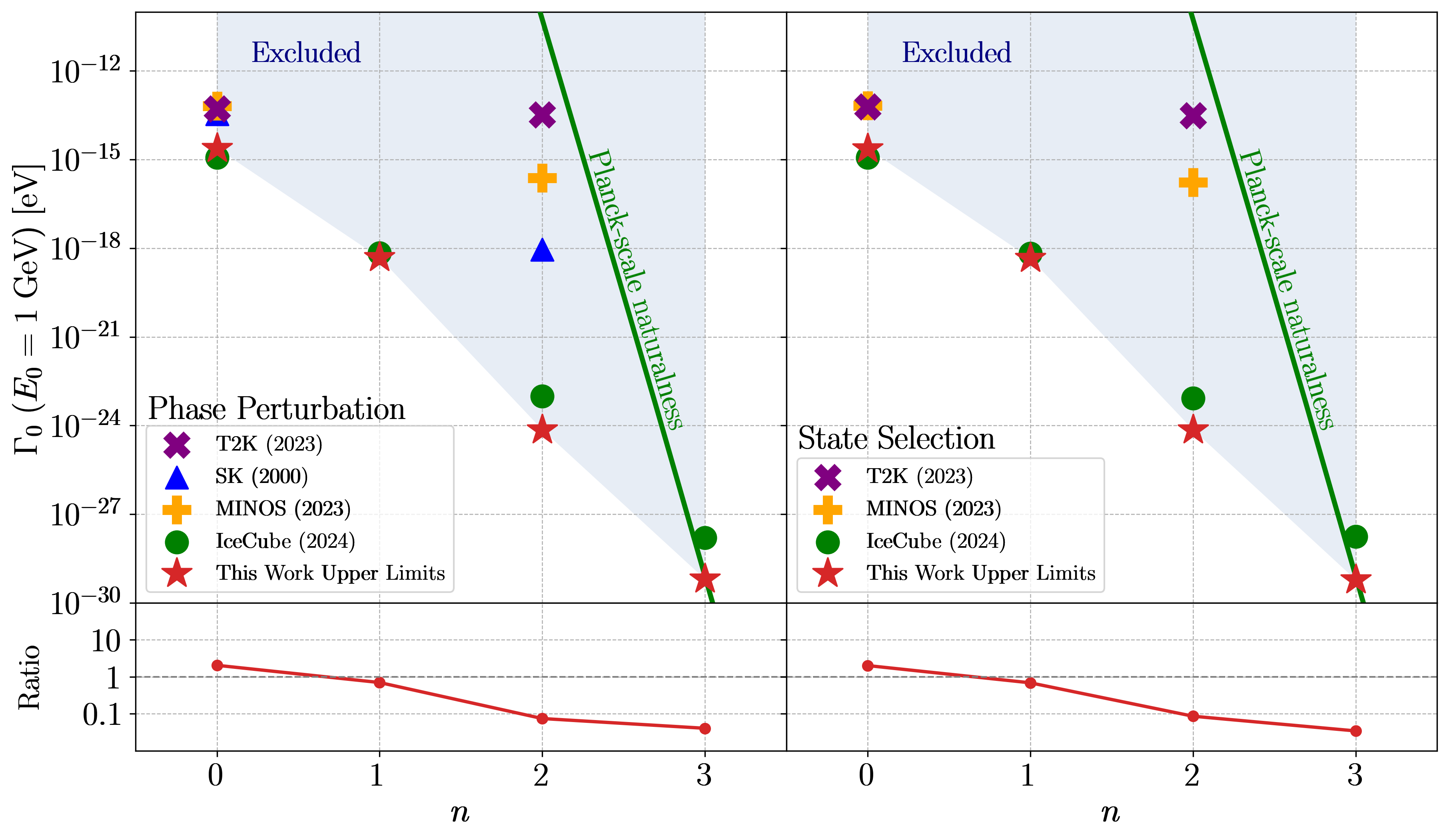}
    \caption{Comparison of past limits for the phase perturbation model (left) and state selection model (right). The bottom plots show the ratio between this work's upper limits and the IceCube 2024 upper limits. Note that $E_0=1\,\text{GeV}$ is used for simpler comparison with past results. The red star markers represent the results from this analysis, and the shading represents the excluded region on the basis of this analysis. The green points represent the results of the prior IceCube analysis for this signature~\cite{gp_deco_paper}. The purple, blue, and orange markers correspond to T2K, SK, and MINOS limits, respectively~\cite{T2K_minos, superk_deco}. The green line indicates the natural Planck-suppressed expectation, obtained by requiring that the decoherence rate reaches the Planck mass $M_{\rm Pl}=1.22\times 10^{19}\,\text{GeV}$ when extrapolated to $E=M_{\rm Pl}$.  }
    \label{fig:comparison}
\end{figure}

We compare these upper limits with previous constraints from IceCube and other long-baseline and atmospheric neutrino experiments in \autoref{fig:comparison}. Relative to the previous IceCube analysis~\cite{gp_deco_paper}, the present results improve the constraints for $n=1$ and strengthen them by more than an order of magnitude for $n=2$ and $n=3$. The improvement is largest for the steepest energy scalings. The high-$n$ improvement is driven primarily by the extended energy range, since the highest-energy events carry the largest effect. The smaller improvement at low $n$ instead comes from the increased event count and improved reconstruction, since an energy-independent decoherence term does not become more visible at higher energies.

For the $n=3$ scenario, the limits now extend below the natural Planck-suppressed expectation obtained from simple dimensional estimates. This scale, shown as the green line in \autoref{fig:comparison}, corresponds to requiring the decoherence rate to reach the Planck scale when extrapolated to $E=M_{\rm Pl}$, i.e. $\Gamma_0(n) = E_0^{n}/M_{\rm Pl}^{n-1}$ with $M_{\rm Pl}=1.22\times10^{19}\,\text{GeV}$. More generally, these results substantially reduce the remaining parameter space for positive energy-dependent decoherence models accessible to atmospheric neutrino measurements.

\section{Conclusions}\label{sec3}
We have searched for neutrino decoherence induced by quantum-gravity-motivated effects using 10.7 years of atmospheric muon-neutrino data collected by the IceCube Neutrino Observatory. Across all tested models the data are consistent with standard three-flavor oscillations, and we set $90\%$ confidence-level upper limits on the decoherence strength for energy dependences $n=0, 1, 2, 3$. Compared to previous IceCube constraints, the limits improve by more than an order of magnitude for $n=2$ and $n=3$.

For $n=3$, these are the first constraints to reach below the natural Planck-suppressed expectation~\cite{stuttard_qg, gp_deco_paper}, in which the decoherence rate reaches the Planck scale when extrapolated to $E=M_{\rm Pl}$. Any decoherence effect of this form must therefore be suppressed more strongly than natural Planck scaling predicts, either through an additional suppression mechanism or through a departure from a simple power-law energy dependence.

Overall, these results provide the most stringent constraints to date on this class of neutrino decoherence models.  Extending the search to 100~TeV, an order of magnitude beyond the previous analysis, drives the improvement at high $n$ and substantially narrows the viable parameter space for Planck-scale open-system effects. Future measurements with larger atmospheric neutrino samples, including those anticipated from IceCube\nobreakdash-Gen2~\cite{Gen2_i_ii, Gen2_iii}, will extend the sensitivity to even weaker quantum-gravity-induced decoherence and enable increasingly stringent tests of open-system effects in neutrino oscillations.

\section{Methods}\label{sec4}
The neutrino-VBH interaction signature was previously searched for with 7.6 years of IceCube data~\cite{gp_deco_paper}, but this new analysis has not only a larger livetime but also a higher purity (increased from 99\% to 99.9\%). The updated analysis also has improved treatment of systematics. These include parameters associated with the conventional and non-conventional neutrino fluxes, overall normalization, detector properties, and neutrino attenuation. The conventional atmospheric flux and cosmic ray uncertainties are modeled using the  \texttt{DaemonFlux} model~\cite{daemonflux}, which includes the hadronic production uncertainties (K$^{\pm}_{158G}$, $\pi^{\pm}_{20T}$, K$^{\pm}_{2P}$, $\pi^{\pm}_{2P}$, p$_{2P}$, n$_{2P}$) and cosmic-ray spectrum uncertainties (GSF$_{1-6}$), and the neutrino and antineutrino fluxes follow the ratio it predicts. The non-conventional (astrophysical) flux is based on IceCube's recent measurement of medium energy starting events~\cite{mese}, specifically using the best fit high energy spectral index and the spectral break energy, while retaining broader priors for the low-energy spectral index and spectral normalization. This choice of prior allows us to exploit additional sensitivity in regions where multiple IceCube datasets are in agreement, while allowing for deviations from a given dataset. This component is taken with equal neutrino and antineutrino fluxes and equal flavor composition, the asymptotic limit of propagation over cosmological baselines, and is propagated through the same decoherence calculation as the atmospheric flux. Assuming a specific source composition instead does not change the results. The analysis systematic uncertainties include the detector DOM efficiency and optical properties of the ice. The bulk ice is modeled using the SnowStorm method~\cite{ice_model}, where Fourier modes parameterize scattering and absorption as a function of ice depth (amplitudes 0--4 and phases 1--4). An additional parameter accounts for the impact of the cloudy refrozen ice in the bore holes in which the DOMs were deployed (forward hole ice). The nominal neutrino-nucleon cross section is taken from the CSMS calculation~\cite{CSMS}. Predictions vary at the level of 5\% depending on the parton distribution functions and target-dependent nuclear effects, and a 10\% uncertainty is adopted to cover this. Attenuation through the Earth above 100~TeV is dominated by interactions with heavy nuclei, where shadowing becomes relevant, and the separate neutrino and antineutrino cross-section parameters ($\nu$ and $\bar{\nu}$ attenuation) model this effect uncorrelated with the overall normalization~\cite{methods&stability}. The decoherence operators are taken to be independent of the matter density along the trajectory, while the standard matter potential is included in the Hamiltonian as usual. 

Each uncertainty contribution is treated as a continuous nuisance parameter in a likelihood maximization analysis, comparing the best-fit likelihood at many values of $\Gamma_0$ to establish a 90\% CL upper limit on $\Gamma_0$ for each power-law index $n$ and decoherence model. The priors and ranges of these parameters can be seen in \autoref{tab:systematics}. This new analysis also separates starting vs. throughgoing tracks and has an improved selection for final-level cuts. 

The likelihood test statistic accounts for both data and Monte Carlo statistical uncertainties, following Ref.~\cite{carlos_frequentist}. We constrain $\Gamma_0$ for $n=0, 1, 2, 3$ using events separated into starting and through-going which are each binned logarithmically into 24 bins in reconstructed muon energy, $\log(E^\mu_{\text{reco}})$, where $E^\mu_{\text{reco}}\in[500\,\text{GeV}, 100\,\text{TeV}]$ and uniformly in 20 bins in zenith angle, $\cos(\theta^\mu_{\text{reco}})$, where $\cos(\theta^\mu_{\text{reco}})\in[-1.0, 0.0]$. 

The robustness of the analysis against systematic uncertainties was further validated through dedicated studies verifying the recovery of injected decoherence signals and the stability of the results under variations of the systematic uncertainty model. IceCube analyses are conducted in a blind manner to avoid experimenter bias. Prior to unblinding, we applied a predefined suite of validation tests based on nuisance-parameter pulls, ensemble consistency, and goodness-of-fit metrics. The resulting goodness-of-fit p-values ranged from $4.02\%$ to $5.40\%$, consistent with expected statistical fluctuations near the null hypothesis, and we therefore proceeded with unblinding.

\begin{table}[ht!]
\footnotesize
\setlength{\tabcolsep}{4pt}
\centering
\begin{tabular}{l|c|c|c|c}
Nuisance parameter & Central value & 1$\sigma$ width of prior & Allowed range & Best-fit pull ($\sigma$) \\
\hline
\hline
\multicolumn{5}{l}{Overall normalization} \\
\hline
Normalization & 1.0 & 0.2 & [0.1, 3.0] & $-0.03$ \\
\hline
\multicolumn{5}{l}{Local response of DOMs} \\
\hline
DOM efficiency & 1.27 & 0.123 & [1.234, 1.346] & $+0.02$ \\
Forward hole ice & $-1.0$ & 10.0 & [$-5.35$, 1.85] & $+0.28$ \\
\hline
\multicolumn{5}{l}{Bulk ice} \\
\hline
Amplitude 0 & 0.0 & 1.0 & [$-3.0$, 3.0] & $+0.72$ \\
Amplitude 1 & 0.0 & 1.0 & [$-3.0$, 3.0] & $+1.28$ \\
Amplitude 2 & 0.0 & 1.0 & [$-3.0$, 3.0] & $+1.30$ \\
Amplitude 3 & 0.0 & 1.0 & [$-3.0$, 3.0] & $+0.72$ \\
Amplitude 4 & 0.0 & 1.0 & [$-3.0$, 3.0] & $+1.09$ \\
Phase 1 & 0.0 & 1.0 & [$-3.0$, 3.0] & $-1.67$ \\
Phase 2 & 0.0 & 1.0 & [$-3.0$, 3.0] & $-0.57$ \\
Phase 3 & 0.0 & 1.0 & [$-3.0$, 3.0] & $-0.20$ \\
Phase 4 & 0.0 & 1.0 & [$-3.0$, 3.0] & $+0.22$ \\
\hline
\multicolumn{5}{l}{Conventional flux} \\
\hline
Zenith correction & 0.0 & 1.0 & [$-3.0$, 3.0] & $-0.17$ \\
Kaon energy loss ($\sigma_{\text{K-Air}}$) & 0.0 & 1.0 & [$-3.0$, 3.0] & $+0.46$ \\
K$_{158G}^{+}$ & 0.0 & 1.0 & [$-2.0$, 2.0] & $+1.01$ \\
K$_{158G}^{-}$ & 0.0 & 1.0 & [$-2.0$, 2.0] & $+0.28$ \\
$\pi_{20T}^{+}$ & 0.0 & 1.0 & [$-2.0$, 2.0] & $+0.00$ \\
$\pi_{20T}^{-}$ & 0.0 & 1.0 & [$-2.0$, 2.0] & $+0.10$ \\
K$_{2P}^{+}$ & 0.0 & 1.0 & [$-2.0$, 2.0] & $+0.18$ \\
K$_{2P}^{-}$ & 0.0 & 1.0 & [$-1.5$, 2.0] & $+0.17$ \\
$\pi_{2P}^{+}$ & 0.0 & 1.0 & [$-2.0$, 2.0] & $-1.61$ \\
$\pi_{2P}^{-}$ & 0.0 & 1.0 & [$-2.0$, 2.0] & $-1.03$ \\
p$_{2P}$ & 0.0 & 1.0 & [$-2.0$, 2.0] & $-0.28$ \\
n$_{2P}$ & 0.0 & 1.0 & [$-2.0$, 2.0] & $-0.20$ \\
GSF$_{1}$ & 0.0 & 1.0 & [$-4.0$, 4.0] & $-0.16$ \\
GSF$_{2}$ & 0.0 & 1.0 & [$-4.0$, 4.0] & $-0.11$ \\
GSF$_{3}$ & 0.0 & 1.0 & [$-4.0$, 4.0] & $+0.00$ \\
GSF$_{4}$ & 0.0 & 1.0 & [$-4.0$, 4.0] & $-0.09$ \\
GSF$_{5}$ & 0.0 & 1.0 & [$-4.0$, 4.0] & $+1.27$ \\
GSF$_{6}$ & 0.0 & 1.0 & [$-4.0$, 4.0] & $-0.85$ \\
\hline
\multicolumn{5}{l}{Non-conventional flux} \\
\hline
\shortstack[l]{Astrophysical normalization ($\Phi^{\text{HE}}$)\\$[10^{-18}\,$GeV$^{-1}$sr$^{-1}$s$^{-1}$cm$^{-2}]$} & 0.787 & 0.36 & [0.0, 3.0] & $+0.29$ \\
$\log_{10}$ of pivot energy, $E^{\text{HE}}_{\text{break}}$/GeV & 4.52 & 0.11 & [4.0, 6.0] & $-0.95$ \\
$\Delta\gamma^{\text{HE}}_{1}$, tilt from $-2.5$ & 0.0 & 0.36 & [$-2.0$, 2.0] & $+2.39$ \\
$\Delta\gamma^{\text{HE}}_{2}$, tilt from $-2.5$ & 0.34 & 0.11 & [$-2.0$, 2.0] & $-0.98$ \\
\hline
\multicolumn{5}{l}{Neutrino attenuation} \\
\hline
$\nu$ attenuation & 1.0 & 0.1 & [0.824, 1.176] & $+0.32$ \\
$\bar{\nu}$ attenuation & 1.0 & 0.1 & [0.824, 1.176] & $-0.05$ \\
\end{tabular}
\caption{Systematic parameters considered in this analysis, with their central values, prior widths, allowed ranges, and best-fit pulls for the $n=2$ phase perturbation model. All parameters are defined in Ref.~\cite{methods&stability}. The hadronic production and cosmic-ray spectrum parameters are those of the DaemonFlux model~\cite{daemonflux}, and the bulk ice amplitudes and phases are Fourier coefficients of the SnowStorm ice model~\cite{ice_model}.}
\label{tab:systematics}
\end{table}
\backmatter

\begin{figure}
    \centering
    \includegraphics[width=0.6\linewidth]{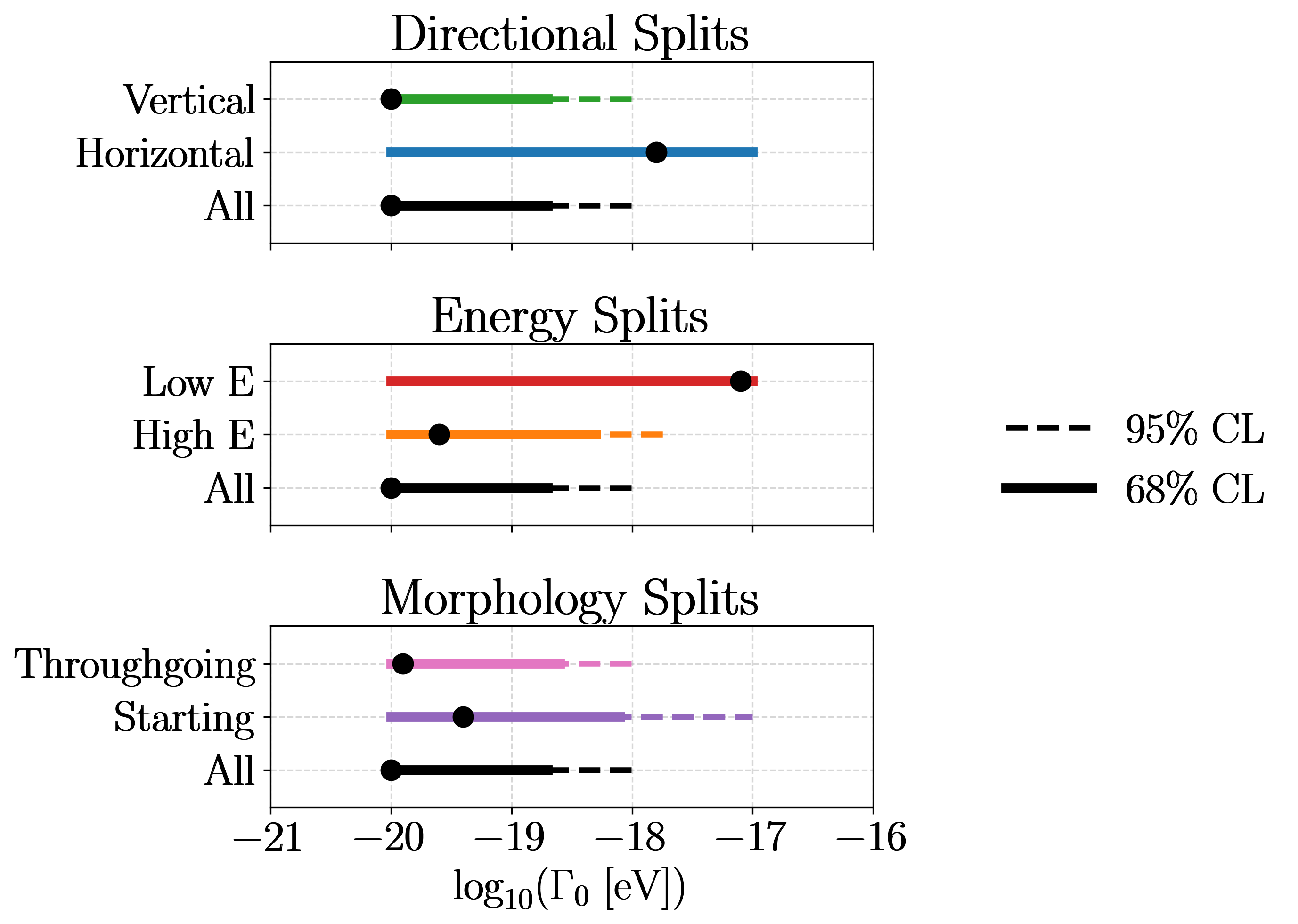}
    \caption{Split sample fit for $n=2$, phase perturbation. Each subplot corresponds to a reconstructed variable used to split the data into two sub-samples. The solid line shows the 68\% CL and the dashed line the 95\% CL for each fit, while the circles denote the best-fit points. The black line shows the overall fit result, and the colored lines show the split-sample fit results. From top to bottom: horizontal vs. vertical (zenith), starting vs. throughgoing, and low vs. high energy.}
    \label{fig:split}
\end{figure}

\bmhead{Supplementary information}
We performed a post-unblinding consistency check for each model by splitting the data into independent subsamples based on zenith (horizontal vs. vertical), energy (low vs. high), and topology (throughgoing vs. starting tracks). An example is shown in \autoref{fig:split}. For each split, we plot the one-dimensional likelihood profile. The solid and dashed contours indicate the 68\% and 95\% confidence intervals, respectively. If there exists a single parameter value that lies within the 95\% confidence interval of the overall fit and each subsample fit on a given subplot, the split results are consistent within $2\sigma$. As shown in~\autoref{fig:split}, this criterion is satisfied for every split. The same is true for all decoherence models, providing an additional validation of the data--Monte Carlo agreement across the analysis phase space.

We have also included a table of our limits compared against those of prior experiments in \autoref{tab:comparison}.
\begin{table*}[htbp]
\centering
\setlength{\tabcolsep}{5pt}
\renewcommand{\arraystretch}{1.15}
\resizebox{\textwidth}{!}{%
\begin{tabular}{llcccc}
\toprule
\textbf{Experiment} & \textbf{Model} & \textbf{$n=0$} & \textbf{$n=1$} & \textbf{$n=2$} & \textbf{$n=3$} \\
\midrule
\multirow{2}{*}{IceCube (this work)}
& State Selection
& $2.36 \times 10^{-15}$
& $4.58 \times 10^{-19}$
& $7.26 \times 10^{-25}$
& $6.05 \times 10^{-30}$ \\
& Phase Perturbation
& $2.41 \times 10^{-15}$
& $4.80 \times 10^{-19}$
& $7.26 \times 10^{-25}$
& $6.37 \times 10^{-30}$ \\
\midrule
\multirow{2}{*}{IceCube (Nature Physics 2024)}
& State Selection
& $1.17 \times 10^{-15}$
& $6.67 \times 10^{-19}$
& $9.48 \times 10^{-24}$
& $1.77 \times 10^{-28}$ \\
& Phase Perturbation
& $1.18 \times 10^{-15}$
& $6.89 \times 10^{-19}$
& $9.80 \times 10^{-24}$
& $1.58 \times 10^{-28}$ \\
\midrule
\multirow{2}{*}{MINOS (JHEP 10 (2023) 035)}
& State Selection
& $6.8 \times 10^{-14}$
& --
& $1.7 \times 10^{-16}$
& -- \\
& Phase Perturbation
& $6.5 \times 10^{-14}$
& --
& $2.4 \times 10^{-16}$
& -- \\
\midrule
\multirow{2}{*}{T2K (JHEP 10 (2023) 035)}
& State Selection
& $6.2 \times 10^{-14}$
& --
& $3.1 \times 10^{-14}$
& -- \\
& Phase Perturbation
& $5.2 \times 10^{-14}$
& --
& $3.3 \times 10^{-14}$
& -- \\
\midrule
Super-K (PRL 85, 1166)
& Phase Perturbation
& $3.5 \times 10^{-14}$
& --
& $0.9 \times 10^{-18}$
& -- \\
\bottomrule
\end{tabular}%
}
\caption{90\% C.L. upper limits on $\Gamma_0$ (eV) from our analysis compared with other searches, using $E_0 = 1~\mathrm{GeV}$.}
\label{tab:comparison}
\end{table*}

\bmhead{Acknowledgements}
The IceCube collaboration acknowledges the substantial contributions to this manuscript from the Harvard University, Massachusetts Institute of Technology, and University of Delaware groups. 
The authors gratefully acknowledge the support from the following agencies and institutions: USA – U.S. National Science Foundation-Office of Polar Programs, U.S. National Science Foundation-Physics Division, U.S. National Science Foundation-EPSCoR, U.S. National Science Foundation-Office of Advanced Cyberinfrastructure, Wisconsin Alumni Research Foundation, Center for High Throughput Computing (CHTC) at the University of Wisconsin–Madison, Open Science Grid (OSG), Partnership to Advance Throughput Computing (PATh), Advanced Cyberinfrastructure Coordination Ecosystem: Services \& Support (ACCESS), Frontera and Ranch computing project at the Texas Advanced Computing Center, U.S. Department of Energy-National Energy Research Scientific Computing Center, Particle astrophysics research computing center at the University of Maryland, Michigan State University, Astroparticle physics computational facility at Marquette University, NVIDIA Corporation, Google Cloud Platform, U.S. Department of Energy, Office of Science, Office of Advanced Scientific Computing Research, and Department of Energy Computational Science Graduate Fellowship; Belgium – Funds for Scientific Research (FRS-FNRS and FWO), FWO Odysseus and Big Science programmes, and Belgian Federal Science Policy Office (Belspo); Germany – Bundesministerium für Forschung, Technologie und Raumfahrt (BMFTR), Deutsche Forschungsgemeinschaft (DFG), Helmholtz Alliance for Astroparticle Physics (HAP), Initiative and Networking Fund of the Helmholtz Association, Deutsches Elektronen Synchrotron (DESY), and High Performance Computing cluster of the RWTH Aachen; Sweden – Swedish Research Council, Swedish Polar Research Secretariat, Swedish National Infrastructure for Computing (SNIC), and Knut and Alice Wallenberg Foundation; European Union – EGI Advanced Computing for research; Australia – Australian Research Council; Canada – Natural Sciences and Engineering Research Council of Canada, Calcul Québec, Compute Ontario, Canada Foundation for Innovation, WestGrid, and Digital Research Alliance of Canada; Denmark – Villum Fonden, Carlsberg Foundation, and European Commission; New Zealand – Marsden Fund; Japan – Japan Society for Promotion of Science (JSPS), Ministry of Education, Culture, Sports, Science and Technology (MEXT), and Institute for Global Prominent Research (IGPR) of Chiba University; Korea – National Research Foundation of Korea (NRF); Switzerland – Swiss National Science Foundation (SNSF). 
\bibliography{sn-bibliography}

\end{document}